\documentclass[sigconf,screen]{acmart}

\AtBeginDocument{ }

\usepackage{amsmath,amssymb,amsfonts}
\usepackage{algorithmic}
\usepackage{graphicx}
\usepackage{textcomp}
\usepackage{xcolor}

\usepackage{hyperref}
\usepackage{makecell}
\usepackage{multirow} 
\usepackage[most]{tcolorbox}
\usepackage{subcaption}
\usepackage{balance}
\usepackage{enumitem}

\usepackage{pifont}

\definecolor{darkgreen}{RGB}{0,120,0}
\newcommand{\redcross}{\textcolor{red}{\ding{55}}}
\newcommand{\greencheck}{\textcolor{darkgreen}{\ding{51}}}

\newtcolorbox{importantBox}{
    colback=gray!10!white,
    boxsep=0.5pt,left=0.5pt,right=0.5pt,top=1pt,bottom=1pt}

\usepackage{amsthm}
\theoremstyle{remark}
\newtheorem*{example}{Example}

\definecolor{codegreen}{rgb}{0,0.6,0}
\definecolor{codegray}{rgb}{0.5,0.5,0.5}
\definecolor{backcolour}{rgb}{0.95,0.95,0.92}
\lstdefinestyle{mystyletext}{
    language=Python,
    backgroundcolor=\color{backcolour},   
    commentstyle=\color{codegreen},
    numberstyle=\tiny\color{codegray},
    basicstyle=\ttfamily\small%
    \lst@ifdisplaystyle\scriptsize\fi,
    breakatwhitespace=false,         
    breaklines=true,                 
    captionpos=b,                    
    keepspaces=true,                 
    showspaces=false,                
    showstringspaces=false,
    showtabs=false,                  
    tabsize=2,
    xleftmargin=2.5mm
}

\begin{document}
\sloppy
\title{JunoBench: A Benchmark Dataset of Crashes in Python Machine Learning Jupyter Notebooks}

\author{Yiran Wang}
\orcid{0009-0007-4613-8960}
\affiliation{%
  \institution{Linköping University}
  \city{Linköping}
  \country{Sweden}
}
\email{yiran.wang@liu.se}

\author{José Antonio Hernández López}
\orcid{0000-0003-2439-2136}
\affiliation{%
  \institution{University of Murcia}
  \city{Murcia}
  \country{Spain}
}
\email{joseantonio.hernandez6@um.es}

\author{Ulf Nilsson}
\orcid{0000-0003-0269-9268}
\affiliation{%
  \institution{Linköping University}
  \city{Linköping}
  \country{Sweden}
}
\email{ulf.nilsson@liu.se}

\author{Dániel Varró}
\orcid{0000-0002-8790-252X}
\affiliation{%
  \institution{Linköping University}
  \city{Linköping}
  \country{Sweden}
}
\email{daniel.varro@liu.se}

\begin{abstract}
Jupyter notebooks are widely used for machine learning (ML) prototyping and experimentation, yet debugging support for notebook-based ML development remains limited, partly due to the lack of realistic and executable bug benchmarks. We introduce JunoBench, to our knowledge the first executable benchmark of real-world crashes in Python-based ML notebooks. 
JunoBench contains 111 curated and reproducible crashes from public Kaggle notebooks, each paired with a verified fix. The benchmark covers widely used ML libraries (e.g., TensorFlow/Keras, PyTorch, and Scikit-learn) as well as notebook-specific failures such as out-of-order execution. To ensure reproducibility and ease of evaluation, JunoBench provides a unified execution environment that reliably reproduces all crashes.
In addition, each crash is accompanied by human-validated annotations, including library cause, crash type, root cause, ML pipeline stage, and natural-language diagnostic summaries. By combining realistic crashes, verified fixes, structured labels, and reproducible execution, JunoBench enables systematic evaluation of crash detection, diagnosis, and automated repair techniques for notebook-based ML development.
\end{abstract}



\maketitle

\section{Introduction}
Machine learning (ML) and deep learning (DL) are increasingly used in modern software programs, with Jupyter notebooks emerging as a dominant development environment for Python-based ML prototyping~\cite{nasahecc2024, Grotov22notebook} (here referred to as \emph{ML notebooks}). Notebooks offer a flexible, interactive interface that supports incremental coding, intermediate feedback, and custom execution order for notebook cells. These features make notebooks well suited for exploratory ML development, but also introduce unique debugging challenges~\cite{Grotov22notebook}. When combined with the inherent complexity of data manipulation and the intensive use of advanced ML and DL libraries, these factors further increase the likelihood of errors during ML prototyping~\cite{islamComprehensiveStudyDeep2019}. 

Existing tools offer limited support for identifying ML bugs, especially in notebook environments~\cite{desantanaBugAnalysisJupyter2024}. This limitation is partly due to the lack of benchmark datasets for systematic evaluation.
Benchmarks are crucial for advancing automatic debugging research by providing standardized datasets of real-world bugs for reproducible evaluation~\cite{Kistowski2015benchmark}. While prior work~\cite{morovatiBugsMachineLearningbased2023, liangGDefect4DLDatasetGeneral2022, kimDenchmarkBugBenchmark2021, zhangEmpiricalStudyTensorFlow2018, wardatDeepLocalizeFaultLocalization2021} focus on ML bugs in Python scripts (\texttt{.py} files), they do not account for the unique semantics of notebooks (\texttt{.ipynb}) such as persistent execution state and out-of-order execution. 

Furthermore, existing ML bug benchmarks (e.g., \cite{morovatiBugsMachineLearningbased2023}) often mix heterogeneous bug symptoms such as performance degradation, incorrect functionality, and crashes. 
Among these, crashes represent the most disruptive symptom, as they terminate program execution, and are the most common in ML programs~\cite{islamComprehensiveStudyDeep2019, morovatiBugsMachineLearningbased2023, desantanaBugAnalysisJupyter2024}. 
Crashes are especially suitable for evaluating debugging techniques because they produce observable and reproducible errors that can be used to assess localization and repair effectiveness.
In the notebook setting, crashes are particularly important, as a crashing cell can alter the notebook’s execution state and lead to unexpected behavior in subsequent cells.

Designing a benchmark for ML notebook crashes requires real-world Jupyter notebooks (i.e., \texttt{.ipynb} artifacts), reproducible execution environments, and validated fixes to support reliable tool evaluation. To meet this need, we present \emph{JunoBench}, an executable benchmark of crashes in real-world Python ML notebooks.
JunoBench supports research on evaluation and development of debugging tools for notebook-based ML development by providing:

\begin{itemize}[leftmargin=*]
    \item 111 curated and reproducible real-world crashes from Python ML notebooks, each paired with a verifiable fix;
    \item Coverage of major ML libraries (e.g., \textit{TensorFlow/Keras}, \textit{PyTorch}, \textit{Scikit-learn}, \textit{Pandas}, \textit{NumPy}, \textit{Matplotlib}, \textit{Seaborn}) as well as notebook-specific failures such as out-of-order execution;
    \item Human-validated crash categorizations (i.e., {library cause}, {crash type}, {root cause}, and {ML pipeline stage}) and ground-truth labels for crash detection and diagnosis;
    \item A unified execution environment (i.e., a Docker image) that enables consistent reproduction across all benchmark cases.
\end{itemize}

To the best of our knowledge, JunoBench is the first benchmark that provides real-world ML notebook crashes, verified fixes, and a unified execution environment for reproducible evaluation of crash identification and debugging tools in the context of ML notebooks.
\section{Methodology}
\begin{figure}
    \centering
    \includegraphics[width=\columnwidth]{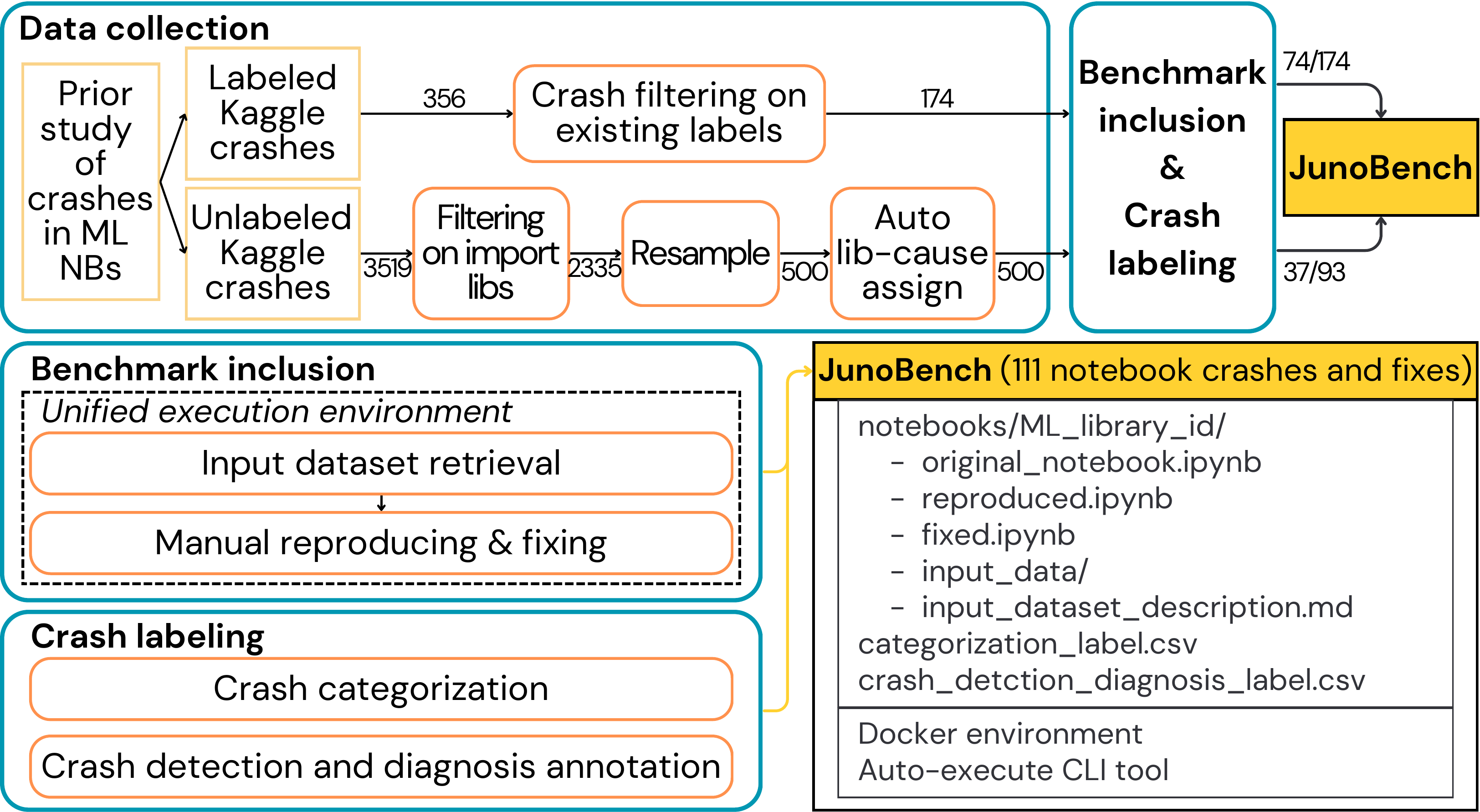} 
    \Description{A plot showing the benchmark construction process.}
    \caption{\small{Overview of the benchmark construction process.}}
    \label{fig: method}
    \vspace{-0.4cm}
\end{figure}

This section outlines our methodology for constructing JunoBench (see~\autoref{fig: method}), including data collection, benchmark inclusion, crash labling, and how we ensure reproducibility and usability. 

\subsection{Data collection}
\subsubsection{Data sources}
\label{subsubsec: method_data_source}
The crashing notebooks in JunoBench are derived from the dataset released in our prior empirical study on public Python ML notebook crashes~\cite{wang2025machinelearningnotebookscrash}. That study analyzed 64,031 notebooks (61,342 from GitHub and 2,689 from Kaggle)
containing crashes. Among these, 746 crashes were manually labeled, while the remaining crashes were provided without manual labels. The released dataset includes detailed annotations for the labeled subset, including the \emph{root cause}, \emph{crash type}, and the ML library used at the point of failure (i.e., the \emph{library cause} label). The labeling procedure of~\cite{wang2025machinelearningnotebookscrash} is revisited in detail in~\autoref{subsubsec: method_categorization}.

\subsubsection{Crash filtering}
First, we filter crashes from the pool of \emph{labeled notebooks} based on the following criteria.
(1) First, we focus exclusively on \emph{Kaggle} notebooks (and exclude GitHub notebooks). Kaggle notebooks typically contain publicly accessible datasets and provide a more controlled execution environment, which facilitates crash reproduction and benchmark reuse.
(2) Then, we include crashes that are (i) related to ML libraries or (ii) caused by notebook-specific issues such as out-of-order execution. This choice aligns with our goal of constructing a benchmark for crash debugging in ML development within Jupyter notebooks, capturing both ML-specific failures and errors arising from incorrect notebook usage.
(3) Finally, we exclude crashes that are hardware- or environment-dependent, including those caused by insufficient computational resources, GPU requirements, incompatible library versions, or bugs inside third-party libraries. Such crashes are difficult to reproduce reliably because the notebooks typically lack detailed environment or hardware specifications, hindering consistent evaluation across different systems.

The prior study~\cite{wang2025machinelearningnotebookscrash} labeled 746 crashing notebooks, of which 356 are from Kaggle. We start from these 356 \emph{labeled} Kaggle crashes and select those that satisfy the inclusion criteria above, based on their annotated \emph{root cause} and \emph{library cause}. After filtering, 174 candidate crashes remain for benchmark construction.

\subsubsection{Resampling}
The filtered labeled crashes are unevenly distributed across ML libraries. Rather than labeling the entire remaining unlabeled Kaggle pool, which would require substantial manual effort and is beyond the scope of this benchmark, we aim to construct a benchmark dataset with balanced coverage across commonly used ML libraries and suitable for controlled evaluation.

The \emph{unlabeled} Kaggle pool of~\cite{wang2025machinelearningnotebookscrash} contains 3,519 crashes (3,875 in total when including the 356 labeled crashes). Because the unlabeled crashes do not include \emph{library cause} annotations, we first apply an automatic coarse-grained filtering step that selects notebooks \emph{importing} at least one target ML library used in the filtered set.

This filtering step yields 2,335 candidate notebooks that import at least one of the target ML libraries. Specifically, the candidate notebooks import the following libraries: \textit{NumPy} (2,131, 91.26\%), \textit{Pandas} (1,976, 84.63\%), \textit{Matplotlib} (1,707, 73.10\%), \textit{Scikit-learn} (1,375, 58.89\%), \textit{Seaborn} (931, 39.87\%), \textit{TensorFlow/Keras} (840, 35.97\%), \textit{PyTorch} (536, 22.96\%), \textit{TorchVision} (242, 10.36\%), \textit{LightGBM} (94, 4.03\%), and \textit{Statsmodels} (93, 3.98\%). This step provides an upper bound on potentially relevant notebooks, since importing a library does not necessarily imply that it is directly involved in the crash.

Then we resample an additional 500 crashes from this filtered unlabeled Kaggle pool. 
The non-linear execution order of notebooks necessitates manual inspection to identify the relevant ML libraries involved in the crashing code line. To support the prioritization of notebooks during resampling, we developed a heuristic tool that automatically estimates the \emph{library cause} of each crash by assigning scores to libraries based on their presence in the crashing code cell and in the traceback of the cell output, with higher weights for libraries appearing on the crashing line. When evaluated on the previously labeled crashes, the heuristic achieves 80\% accuracy.
The heuristic is used solely to prioritize candidates for manual inspection and inclusion, enabling us to construct a benchmark with balanced coverage across ML libraries while keeping annotation effort tractable.

\subsection{Benchmark inclusion}

\subsubsection{Input dataset retrieval}
For each candidate notebook, we first retrieve the metadata of the required input datasets using KGTorrent~\cite{quarantaKGTorrentDatasetPython2021} with the Meta Kaggle~\cite{megan2022metakaggle} dataset, including dataset titles, downloading links, and licenses. 

As our focus is on crash reproduction rather than the performance of ML models, the use of full training data is not relevant. 
For input datasets exceeding 100MB, we evaluated whether downsampling preserves the crash behavior without altering code logic. If the same crash is reproduced, we include the downsampled dataset in our benchmark instead of the full dataset. For instance, a 2.5GB image dataset was reduced to 94MB by retaining just 4\% of images per class while still reproducing the crash.

\subsubsection{Manual reproducing and fixing}
In the prior empirical study~\cite{wang2025machinelearningnotebookscrash}, the dataset of notebooks was organized at the crash level, where multiple crashes could originate from the same notebook. 
In JunoBench, we instead design each benchmark instance as an independent notebook containing a single, isolated, and reproducible crash. This ensures that each case is self-contained, simplifying evaluation and comparison across automated debugging tools.

For each notebook, we maintain three versions: (1) the \emph{original} notebook as collected, (2) a \emph{reproduced} version containing minimal adjustments (e.g., the sequence of execution order for crash reproduction, dataset paths, training settings),
and (3) a \emph{fixed} version where the crash is repaired.

A reproduced crash is required to \emph{exactly match} the original Kaggle execution behavior, i.e. (1) it reproduces the same crashing code cell and (2) exhibits the same error output (i.e., exception type, error message, failure localization, and traceback structure). Differences that are unrelated to crash information in the error outputs, such as file paths, environment-specific prefixes, execution counters, or cell numbering, are ignored. Notebooks are excluded from the benchmark if the reproduced execution deviates from the original crash behavior, or the crash could not be reproduced. Hence, all crashes included in JunoBench represent real crashes, but in a more compact environment with less input data.

For reproducing each crash, we execute only the code cells required to trigger the crash, yielding one valid execution sequence that reproduces the original crash (although alternative sequences may exist). This sequence is reflected in the cell execution counts, which are preserved in the fixed version for comparability.

Fixes are produced manually with minimal code edits based on inspection of the reproduced failure behavior, including the error message, traceback, and failing code context. All fixes eliminate the crash while preserving the notebook’s original intent. To improve transparency and traceability, fix locations are explicitly marked in the fixed version using standardized inline comments (e.g., \textit{\# fix - - -}), enabling direct comparison between reproduced and fixed versions.
We analyze and group the applied fixes retrospectively into a small number of recurring fix patterns, which can be related to the root causes and crash types introduced later.

The applied fixes fall into the following recurring action-level categories, including:
\begin{enumerate}[leftmargin=*]
    \item \emph{Correcting API misuse}, such as adapting inputs or arguments to satisfy library API contracts (e.g., data shape or type constraints, valid argument values), correct attribute or method usage, or updating incompatible API calls;
    \item \emph{Repairing data-related issues}, including correcting data preprocessing steps and resolving mismatches in data shape, type, schema, or value properties;
    \item \emph{Fixing implementation errors}, through localized code edits, such as correcting faulty logic, adding missing conditions, or completing omitted computation steps;
    \item \emph{Adjusting model configuration or structures}, including fixing invalid model initialization arguments or inconsistent model structures;
    \item \emph{Resolving notebook-specific state issues}, such as reordering or re-executing cells to address inconsistencies caused by non-linear notebook execution.
\end{enumerate}

\begin{example}
\autoref{lst: fix_example} shows a representative example of a \emph{category (1)} fix addressing an API misuse that leads to a tensor shape mismatch. In this case, the ground-truth labels are one-hot encoded, whereas the loss API (\texttt{nn.CrossEntropyLoss}) expects class indices in a one-dimensional tensor. The fix converts the labels using \texttt{argmax} to obtain class indices before computing the loss, thereby resolving the mismatch without modifying the model logic. The complete set of fixes and their corresponding reproduced-fixed notebook pairs are released as part of the dataset~\cite{junobenchrepo}.
\begin{lstlisting}[style=mystyletext, caption=\small{Example of a minimal manual fix in JunoBench addressing an API misuse crash caused by a tensor shape mismatch (\texttt{RuntimeError: 0D or 1D target tensor expected, multi-target not supported}).}, label={lst: fix_example}]
- loss_ = loss(output, label).to(device)
+ # fix --- label is one-hot encoded, but nn.CrossEntropyLoss expects 1D targets
+ label = label.argmax(dim=1)
+ loss_ = loss(output, label).to(device)
\end{lstlisting}

\end{example}

We aim to construct a balanced benchmark across ML libraries with inclusion of notebook-specific issues.
For the 174 previously labeled candidates, we successfully reproduce and fix 74 crashes.
We then continue with the 500 unlabeled candidates, prioritizing reproduction based on the automatically inferred \emph{library cause}.
To maintain a balanced coverage of ML libraries, we monitor library distribution throughout the process and halt reproduction once sufficient coverage is achieved. 
From the unlabeled set, 93 crashes are manually examined, and 37 are successfully reproduced and fixed, while others are excluded primarily due to missing or inaccessible datasets. 

In total, JunoBench includes 111 reproducible crashes with verified fixes, emerging from the multi-stage benchmark inclusion pipeline described above. It covers both notebook-specific execution issues and a balanced set of crashes across common ML libraries (\autoref{fig: lib}).

\subsection{Crash labeling}

\subsubsection{Crash categorization}
\label{subsubsec: method_categorization}
We adapt the crash categorization scheme and annotation protocol from the prior study~\cite{wang2025machinelearningnotebookscrash} to annotate all crashes in JunoBench. 

The labeling process in the prior study~\cite{wang2025machinelearningnotebookscrash} followed a grounded theory methodology~\cite{seaman1999quality, saldana2015coding}. Three annotators were involved. In the first cycle, all three annotators independently labeled an overlapping subset of notebooks to establish a shared coding scheme. In the second cycle, the full set of labeled notebooks was annotated, and each annotation was subsequently reviewed by another annotator (i.e., no annotator reviewed their own labels). Disagreements were discussed and resolved through consensus. The study reports disagreement rates across labeling dimensions (e.g., 9.65\% for root cause and 4.29\% for library cause), and the final released labels reflect the consensus decisions.

In this work, we follow the same grounded theory-based labeling procedure, including the labeling criteria, annotator instructions, and adjudication process from that study, without modification. This ensures consistency and facilitates comparative analyses. 
Each crash is labeled along four dimensions:  (1) \emph{library cause}, (2) \emph{crash type} (e.g., tensor shape mismatch, unsupported broadcast), (3) \emph{root cause} (e.g., API misuse, data confusion), and (4) \emph{ML pipeline stage} (i.e., where in the general ML workflow the crash occurs, see~\autoref{fig: mlpp}).

For the 74 reproducible labeled crashes from prior study~\cite{wang2025machinelearningnotebookscrash}, we revalidate and refine the annotations as the prior study did not re-execute the notebooks or establish and validate fixes for the crashes. Revalidation involves re-executing each crash, inspecting the failing code cell, error message, and traceback, and confirming that the annotated labels are consistent with the observed failure and the applied fix. Refinement is limited to correcting label assignments, primarily for the \emph{root cause} dimension, when the original label is found to be inconsistent with the confirmed cause revealed by the fix, while label definitions and granularity are not changed.
As a result, seven root-cause labels (9.5\%) are corrected. For example, a crash initially labeled as \emph{data confusion} is reclassified as \emph{API misuse} after identifying an incorrect argument value passed to the \texttt{plot\_acf} API in \texttt{statsmodels}. All corrections are documented in our GitHub repository~\cite{wang2025code}.

For the 37 unlabeled crashes, we apply grounded theory methods~\cite{seaman1999quality, saldana2015coding}, using first-cycle coding by one annotator and second-cycle review by a different annotator. 
Inter-annotator agreement is assessed using Cohen’s $\kappa$, computed separately for each labeling dimension. Observed agreement exceeds 0.9 for all dimensions, indicating almost perfect agreement. The resulting $\kappa$ values are 0.9 for \emph{root cause}, 0.93 for \emph{ML pipeline stage}, and reached 1.0 for \emph{library cause} and \emph{crash type}.
Disagreements are resolved through discussion. 
A total of five disagreements are discussed, and full consensus is reached on all final labels.

\subsubsection{Crash detection and diagnosis annotation}
To support automated crash identification and diagnosis, we provide ground-truth annotations at two levels: (1) a \emph{crash detection label}, assigned at the notebook level (\textit{true} for reproduced, \textit{false} for fixed), and (2) a \emph{crash diagnosis label}, describing the cause and localization of each crash with a short natural language explanation. 

Each diagnosis label is initially produced by one annotator based on evidence from the crashing code cells, error outputs, and the corresponding fixed versions, and then independently reviewed by a second annotator for correctness and completeness.
Given that diagnosis labels are descriptive natural language summaries of failure mechanisms, we adopt an expert validation protocol in which the second annotator performs targeted review rather than parallel blind annotation.
During review, the second annotator either accepts the diagnosis or flags issues for discussion. In total, six diagnosis labels require discussion, and all are resolved through consensus between the two annotators.

\subsubsection{Input dataset license}
Most datasets in JunoBench are publicly available and licensed for research use. In four cases where the original datasets (all numeric) are under restrictive license, we replace them with synthetic equivalents by replicating the schema and missing-value structure while generating random values.
Finally, each dataset is accompanied by a metadata file documenting its title, source URL, license, and modifications (e.g., downsampling or synthetic substitution).

\subsection{Benchmark usability}
To enable reliable execution of ML notebook crashes, JunoBench provides a unified execution environment packaged as a Docker image. 
Notebook reproducibility is often sensitive to dependencies and library interactions~\cite{pimentelLargeScaleStudyQuality2019}. 
Providing a unified environment therefore ensures that crashes can be reproduced consistently across all benchmark cases and evaluated under comparable conditions.

Our Docker image is built on top of the official Kaggle CPU Docker image from the Kaggle repository~\cite{kaggle2022dockerrepo}, reflecting a realistic ML development setting. We extend this base image with all dependencies required to execute JunoBench notebooks, as specified in \texttt{requirements.txt} in the JunoBench repository~\cite{junobenchrepo}. To ensure long-term reproducibility independent of upstream image availability, we archive the fully built JunoBench Docker image and provide image digests and build details in the project repository~\cite{wang2025code}.

This one-time setup enables consistent execution across all benchmark cases while reducing per-case configuration overhead. 
To support extensibility, we also provide build artifacts (e.g., Dockerfile, dependency specifications, and documentation) that allow researchers to customize or reconstruct environments when needed.


We additionally provide an execution-level interface inspired by a prior study~\cite{zhuBugsBenchmarksComprehensive2025a}, implemented as a command-line tool.
It automatically runs both reproduced and fixed versions with provided cell execution order, and verifies that the reproduced version crashes while the fixed version executes successfully. 
This interface enables users to reproduce and validate all benchmark cases with a single command.

\section{JunoBench}
\label{sec: JunoBench}
We present the characteristics of JunoBench in four dimensions: \emph{library cause}, \emph{root cause}, \emph{crash type}, and \emph{ML pipeline stage}. 

\begin{figure}[t]
    \centering
    \begin{subfigure}{0.33\linewidth}
        \centering
        \includegraphics[width=\linewidth]{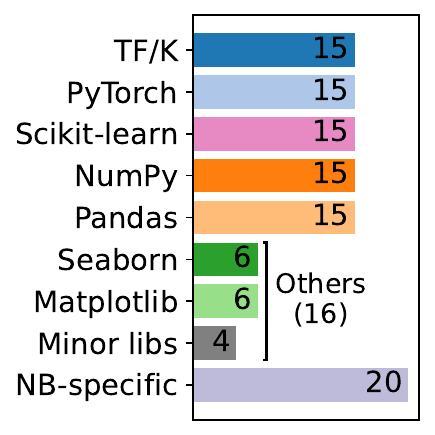} 
        \caption{\small{Library cause}}
        \label{fig: lib}
    \end{subfigure}
    \begin{subfigure}{0.66\linewidth}
        \centering
        \includegraphics[width=\linewidth]{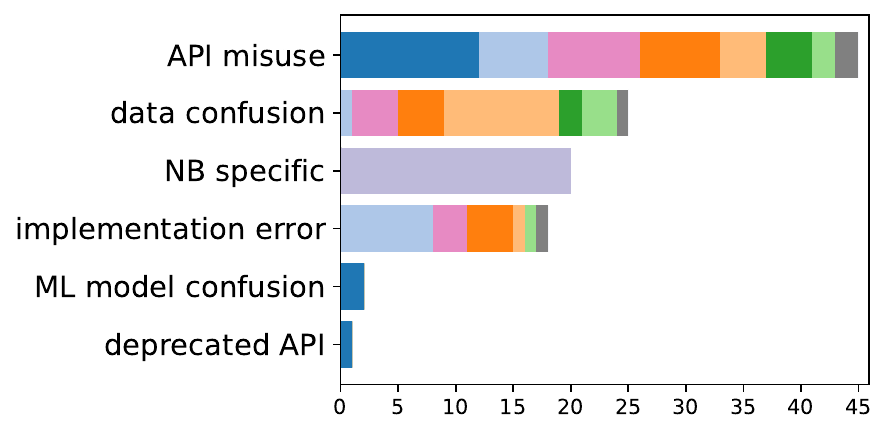} 
        \caption{\small{Root cause}}
        \label{fig: rootcause}
    \end{subfigure}
    \begin{subfigure}{0.39\linewidth}
        \centering
        \includegraphics[width=\linewidth]{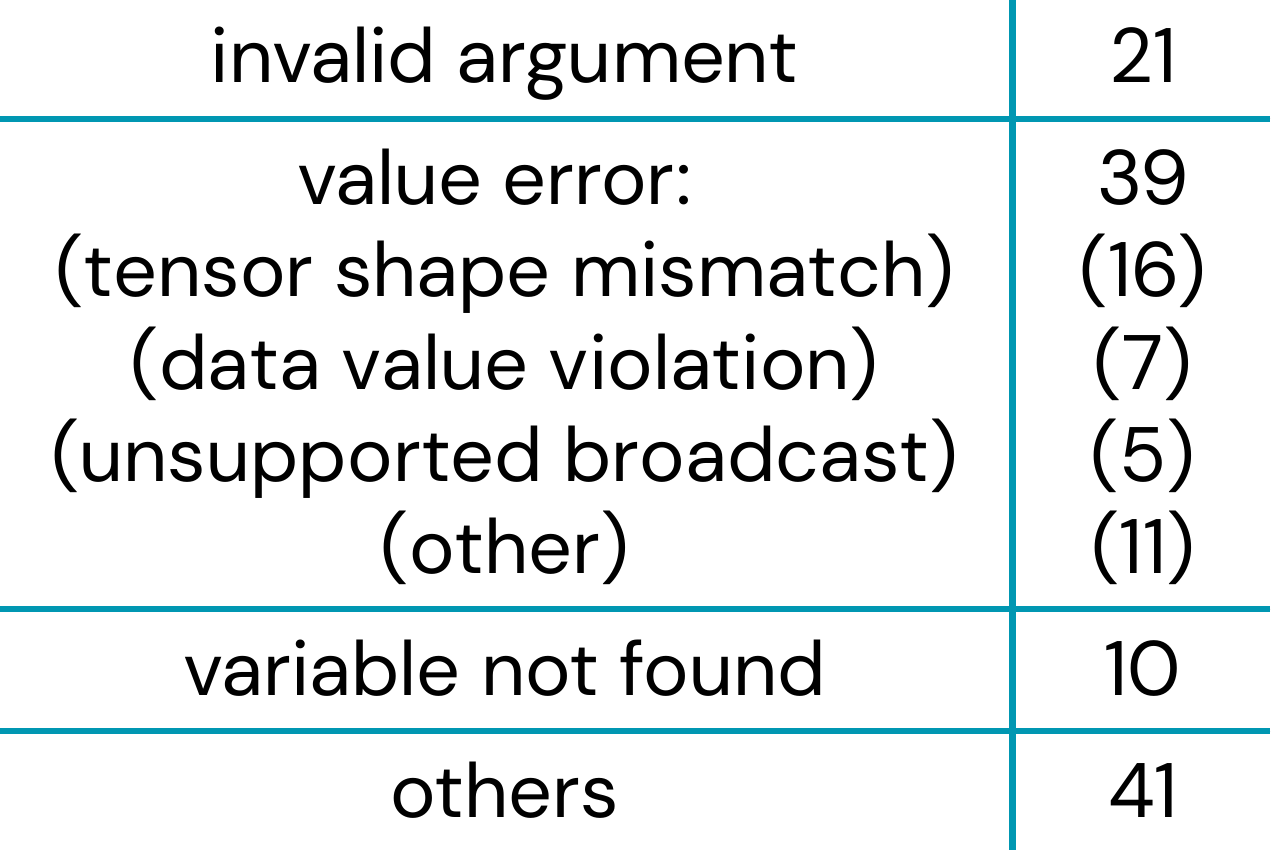} 
        \caption{\small{Crash types}}
        \label{fig: crashtype}
    \end{subfigure}
    \begin{subfigure}{0.6\linewidth}
        \centering
        \includegraphics[width=0.9\linewidth]{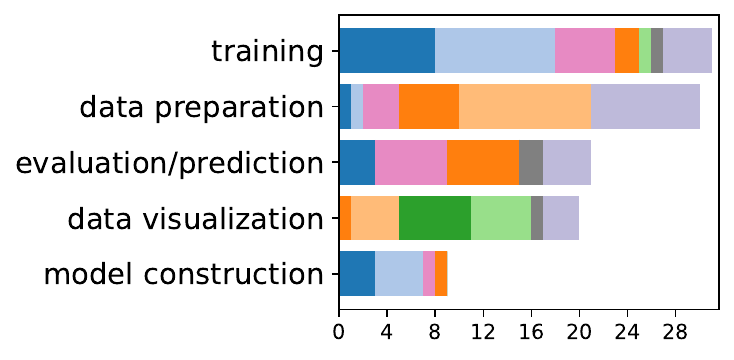} 
        \caption{\small{ML pipeline stage}}
        \label{fig: mlpp}
    \end{subfigure}
    \Description{Plots and tables showing the statistics of JunoBench.}
    \caption{\small{Characteristics of JunoBench. Each bar in (b) and (d) is segmented by libraries (a). “TF/K” stands for “TensorFlow/Keras”. “Minor libs” include Statsmodels(2), TorchVision(1), and LightGBM(1).}}
    \vspace{-0.4cm}
\end{figure}

\paragraph{Library cause}
JunoBench comprises 111 notebook crashes with balanced coverage across popular ML libraries, as well as notebook-specific issues. The distribution is shown in~\autoref{fig: lib}. Specifically, the benchmark includes:
\begin{itemize}[leftmargin=*]
    \item 15 crashes each involving the DL libraries \textit{TensorFlow/Keras} and \textit{PyTorch}, the classical ML library \textit{Scikit-learn}, and the data processing libraries \textit{NumPy} and \textit{Pandas};
    \item 16 crashes involving a mix of ML libraries: 12 related to visualization (6 from \textit{Seaborn} and 6 from \textit{Matplotlib}) and individual cases from \textit{Statsmodels}, \textit{TorchVision}, and \textit{LightGBM};
    \item 20 notebook-specific (i.e., out-of-order cell execution) crashes.
\end{itemize}

This distribution highlights JunoBench’s diverse coverage of challenges in ML notebook development, spanning DL, classical ML, data processing, and visualization libraries, as well as execution order issues unique to interactive notebook environments~\cite{wang2025machinelearningnotebookscrash}.

\paragraph{Root cause}
The root causes of the 111 crashes are shown in~\autoref{fig: rootcause}. In addition to the 20 \emph{notebook-specific} out-of-order execution failures (e.g., re-running a code cell that deletes a column, resulting in an error when the column is referenced again), the majority of remaining crashes fall into three main categories. The most frequent is \emph{API misuse} (45 cases), where developers call library APIs with invalid arguments or unsupported usage patterns. \emph{Data confusion} (25 cases) follows, stemming from misunderstandings of data shape, type, or structure. \emph{Implementation errors} (18 cases) account for logic mistakes such as incorrect variable references or flawed algorithm design. 
JunoBench also includes less common but real-world ML issues. For example, \emph{ML model confusion} (2 cases) occurs when attempting to load a model with an incompatible architecture or from a different framework. 
The rest are caused by \emph{deprecated APIs} that were removed in newer library versions.

\paragraph{Crash type}
Crashes in JunoBench span a variety of types, as summarized in~\autoref{fig: crashtype}. The most frequent is \emph{invalid argument} (21 cases), where function calls violate API constraints due to incorrect or missing parameters. Tensor-related issues such as \emph{tensor shape mismatch} (16) and \emph{unsupported broadcast} (5) are also common, particularly in \textit{TensorFlow/Keras} and \textit{PyTorch} notebooks, where tensor alignment is often complex. \textit{Data value violation} (7) occurs when input values violate API or code assumptions.
Another frequent category is \emph{variable not found} (10), all caused by out-of-order cell execution, where variables defined in unexecuted cells are referenced later.
These crash types capture how developers frequently struggle with runtime dynamics, data integrity, and ML library usage, reflecting ML-specific challenges in notebook development.

\paragraph{ML pipeline stage}
JunoBench captures crashes across key stages of a typical ML pipeline (\autoref{fig: mlpp}). 
Most crashes occur during \emph{model training} (31 cases), primarily involving \textit{TensorFlow/Keras} and \textit{PyTorch}, reflecting the complexity of using DL libraries. 
\emph{Data preparation} follows closely (30), often affected by out-of-order execution and data frame handling errors involving \textit{Pandas}. 
Model \emph{evaluation and prediction} account for 21 crashes, typically involving incorrect metric usage or incompatible model inputs related to \textit{TensorFlow/Keras}, \textit{Scikit-learn}, and \textit{NumPy}. 
\emph{Data visualization} introduces 20 crashes, largely tied to \textit{Matplotlib} and \textit{Seaborn}, but also reflecting upstream data processing issues in \textit{Pandas} and \textit{NumPy}. 
Finally, \emph{model construction} contributes 9 crashes, mostly due to misconfigured models with DL libraries.
\section{Potential research opportunities}
JunoBench enables systematic investigation of crashes in stateful and exploratory ML notebook development. By providing reproducible crashes with validated fixes across multiple ML libraries and notebook-specific errors, structured categorizations (i.e., library cause, crash type, root cause, and ML pipeline stage), ground-truth labels of crash detection and diagnosis, and a unified execution environment, the benchmark supports research that cannot be easily conducted using existing script-based ML bug datasets.

\paragraph{Crash detection, diagnosis, and automated repair}
Notebook crashes frequently depend on interactions across multiple cells and the persistent kernel state accumulated during cell execution~\cite{desantanaBugAnalysisJupyter2024, macke21nbsafety}. Therefore, crashes cannot be understood by inspecting a single code cell in isolation. In practice, developers must reason about the executed cell sequence that leads to the crash and identify which prior cells are responsible for the root cause of the crash. As such crash detection, diagnosis, and repair is a cross-cell challenge.

JunoBench supports research in this setting by providing a sequence of executed cells that reliably reproduces each crash, together with verified fixes and structured annotations. This enables methods that infer cross-cell dependencies, diagnosis or localize the crash-inducing step within a concrete execution trace, and generate repairs that account for notebook kernel state. Therefore, the benchmark facilitates systematic evaluation of such debugging techniques for ML notebooks.

\paragraph{Pipeline-aware and cross-library debugging}
Developers typically build ML/DL programs as pipelines consisting of multiple stages such as data preparation, model training, and evaluation~\cite{zhang20DLjobs}. These stages often involve multiple ML libraries. In notebook environments, crashes are frequently associated with specific pipeline stages and particular libraries~\cite{wang2025machinelearningnotebookscrash}. For example, data transformations performed during the data preparation stage may produce shapes or types that are incompatible with downstream ML models, or mismatched assumptions may arise between preprocessing steps and learning library APIs. In such situations, effective debugging requires understanding not only the crashing line but also the pipeline stage in which the crash occurs and how earlier stage outputs constrain downstream behavior.

JunoBench enables research in pipeline-aware and cross-library debugging by providing categorization labels for each crash with its corresponding ML pipeline stage and involved ML library within realistic notebook workflows. This allows researchers to study stage-specific crash characteristics and design debugging approaches that incorporate pipeline context. The benchmark covers multiple common ML libraries, which further supports evaluation of techniques to see if its performance generalizes across libraries.

\paragraph{Large language model (LLM)-based crash analysis in ML notebooks}
Recent work~\cite{Jahan25empiricalbuglocalDL} has applied LLMs to understand and diagnose crashes in ML code.
In ML notebooks, crashes often depend on multi-cell context, execution state, and library semantics. Therefore, compared with traditional code tasks, crash analysis in notebooks requires LLMs to reason over stateful execution, cross-cell dependencies, and, potentially, program state in the notebook kernel, which is a unique setting to evaluate LLM debugging capabilities.

JunoBench supports research on LLM-based crash analysis by providing human-validated ground-truth for crash detection and diagnosis together with reproducible execution traces and verified fixes. This enables systematic evaluation of LLM reasoning over notebook context and facilitates comparison across LLMs, improvement strategies, and analysis settings. CRANE-LLM~\cite{wang2026cranellm} illustrates this use case by studying LLM-based crash detection and diagnosis, reporting \emph{F1-scores between 57.7\% and 64.6\%} across state-of-the-art models when reasoning over notebook code cells. These results suggest that understanding crashes in a notebook setting remains a challenge and highlight opportunities for improving LLM reasoning in the context of ML notebooks.

\paragraph{Agent-based debugging for notebooks}
Debugging notebooks is an interactive process in which developers iteratively inspect intermediate results, query runtime objects, and decide what to execute next before applying fixes. Therefore, recent research explores LLM-based debugging agents and notebook assistants, which are systems that autonomously perform debugging actions such as inspecting variables, executing cells, and proposing fixes within the notebook environment~\cite{grotov2024nbagent, levin25chatdbg}. These approaches require benchmarks that support step-wise reasoning over execution context and evaluation of action sequences rather than single-shot predictions.

JunoBench enables research in this scenario by providing reproducible notebook crashes together with executable cell sequences, a reliable execution environment with accessible runtime context, and verified fixes. This setup allows evaluation of agent-based debugging approaches that achieve crash diagnosis and repair through performing notebook actions such as cell execution, cell edits, and kernel inspection.

Together, these directions position JunoBench as a major foundation for evaluating notebook reliability in future research.



\section{Related work}

\begin{table*}[ht]
    \footnotesize
    \centering
    \caption{\small{Comparison of prior ML bug datasets and studies across key dimensions: artifact type, crash focus, whether the bugs were reproduced by the authors, whether environment specifications are provided (Env. spec.), and study scope.}}
    \begin{tabular}{c|c|c|c|c|c}
        \toprule
        \thead{Study / Dataset} & \thead{Artifact type} & \thead{Crash-focused} & \thead{Reproduced} & \thead{Env. spec.} & \thead{Scope}\\
        \midrule
        Defect4ML~\cite{morovatiBugsMachineLearningbased2023} 
        &
        Scripts
        &
        Mixed
        &
        \greencheck
        &
        \greencheck (Per-case)
        &
        TensorFlow/Keras\\
        
        \hline
        
        gDefects4DL~\cite{liangGDefect4DLDatasetGeneral2022} 
        & 
        Scripts
        & 
        Mixed
        &
        \greencheck
        &
        \greencheck (Per-case)
        &
        TensorFlow/Keras, PyTorch\\
        
        \hline
        
        Denchmark~\cite{kimDenchmarkBugBenchmark2021}
        & 
        Bug reports
        &
        Mixed
        &
        \redcross
        &
        \redcross
        &
        Broad DL projects\\
        
        \hline
        
        Zhang et al.~\cite{zhangEmpiricalStudyTensorFlow2018}
        & 
        Scripts
        &
        Mixed
        &
        \greencheck
        &
        \redcross
        &
        TensorFlow\\
        
        \hline
        
        Tensfa~\cite{wuTensfaDetectingRepairing2021} 
        & 
        Scripts
        &
        Partial (shape crashes)
        &
        \greencheck
        &
        \redcross
        &
        TensorFlow/Keras\\
        
        \hline

        De Santana et al.~\cite{desantanaBugAnalysisJupyter2024} 
        & 
        Notebooks
        &
        Mixed
        &
        \redcross
        &
        \redcross
        &
        General notebook bug analysis\\
        
        \hline
        
        Wang et al.~\cite{wang2025machinelearningnotebookscrash} 
        & 
        Notebooks
        &
        \greencheck
        &
        \redcross
        &
        \redcross
        &
        ML notebook crash analysis\\
        
        \midrule
        
        \textbf{JunoBench} 
        & 
        \textbf{Notebooks}
        &
        \textbf{\greencheck}
        &
        \textbf{\greencheck}
        &
        \textbf{\greencheck (unified Docker image)}
        &
        \textbf{Multiple ML libraries, notebook issues}\\
        
        \bottomrule
        
    \end{tabular}
    \label{tb: relatedwork}
\end{table*}

As~\autoref{tb: relatedwork} shows, several studies have introduced bug benchmarks in Python-based ML programs. These benchmarks differ along key dimensions such as artifact type (Python scripts vs. Jupyter notebooks), bug symptoms (crashes vs. broader defects), level of reproducibility support (e.g., whether the authors reproduced the bugs, whether environment specifications are provided for external reproduction and execution) and scope.

\paragraph{Script-based ML bug benchmarks}
Early ML bug benchmarks primarily target ML/DL projects implemented as Python scripts.
Defect4ML~\cite{morovatiBugsMachineLearningbased2023} by Morovati et al. contains 100 reproducible bugs collected from \textit{TensorFlow/Keras} projects using sources such as GitHub, Stack Overflow, and prior studies~\cite{zhangEmpiricalStudyTensorFlow2018, wardatDeepLocalizeFaultLocalization2021, NikanjamFaultDetection2021, islamRepairDNN2020}. 
The benchmark includes real faults and executable artifacts, but focuses on \textit{TensorFlow/Keras} programs and mixes crashes with other defect types such as performance regressions. Moreover, the artifacts are program scripts rather than Jupyter notebooks.

Liang et al.~\cite{liangGDefect4DLDatasetGeneral2022} proposed gDefects4DL, a curated dataset of 64 DL bugs drawn from \textit{TensorFlow/Keras} and \textit{PyTorch} projects. 
Similar to Defect4ML, it provides reproducible buggy and fixed program versions, but mixes multiple defect categories and targets script-based projects rather than notebook workflows.

At a larger scale, Kim et al. introduced Denchmark~\cite{kimDenchmarkBugBenchmark2021}, which comprises 4,577 bug reports from DL projects. While offering broad coverage of reported issues, it does not provide reproduction-ready artifacts, making it unsuitable for studies requiring program execution or tool evaluation~\cite{zhuBugsBenchmarksComprehensive2025a}.

Other datasets have emerged as byproducts of empirical studies. For example, Zhang et al.~\cite{zhangEmpiricalStudyTensorFlow2018} reproduced 151 \textit{TensorFlow} bugs in an empirical study, and
Tensfa~\cite{wuTensfaDetectingRepairing2021} focuses on tensor shape mismatches in \textit{TensorFlow/Keras} programs. 
These datasets provide valuable insight into ML fault characteristics but they are specialized either by ML libraries (e.g., \textit{TensorFlow}) or bug type (e.g., tensor shape mismatches) and target Python scripts. Moreover, they are not designed as executable benchmarks with reproducible environments and explicit library dependency specifications, limiting their relevance and reproducibility to modern ML development~\cite{jahangirovaRealFaultsDeep2024}. 

\paragraph{Notebook-based ML bug benchmarks}
In contrast to script-focused ML benchmarks, prior work on Jupyter notebooks has largely focused on general notebook usage~\cite{Grotov22notebook, quarantaKGTorrentDatasetPython2021}, reproducibility~\cite{wangRestoreExecution2021}, or quality analysis~\cite{pimentelLargeScaleStudyQuality2019, quarantaElicitingBestPractices2022}, rather than curated bug benchmarks.
Existing notebook datasets support studies of notebook behavior, best practices, and reproducibility failures, but typically do not provide validated buggy-fixed pairs with executable environments.

One exception is the empirical study by De Santana et al.~\cite{desantanaBugAnalysisJupyter2024}, which analyzes bugs in Jupyter notebook projects. Their work examines general bugs in Jupyter notebooks such as environment setup, kernel, notebook conversion, and implementation errors. However, they did not target bugs in ML code or reproduce the reported bugs, and the released dataset does not include reproducible artifacts, such as validated buggy and fixed notebooks, execution environments, or input data.

A closely related line of work is the empirical crash analysis of ML notebooks by Wang et al.~\cite{wang2025machinelearningnotebookscrash}, which categorizes ML-specific, general Python, and notebook-related issues.
This study characterizes the problem space and provides a taxonomy of crashes, but does not reconstruct execution environments, or release validated buggy-fixed notebook pairs.

\paragraph{Summary}
Existing datasets have several limitations.
(1) Most ML bug benchmarks focus on script-based projects rather than notebook artifacts that dominate interactive ML development.
(2) Many datasets mix bugs with heterogeneous symptoms (e.g., crash, bad performance, incorrect functionality), which complicates evaluation of crash-driven tasks. 
(3) Existing benchmarks are often library-specific (e.g., focus exclusively \textit{TensorFlow/Keras} or \textit{PyTorch}), whereas real-world ML development typically involves multiple ML and data science libraries.
(4) Finally, reproducibility support varies substantially, with limited availability of standardized execution environments that enable consistent evaluation across cases.

To address these limitations, JunoBench introduces an executable benchmark centered on crashes in real-world ML notebooks. The benchmark provides curated buggy-fixed notebook pairs validated through re-execution and a unified execution environment that enables reliable reproduction and comparable evaluation across cases. In addition, JunoBench includes crash categorization labels and diagnosis annotations to support fine-grained evaluation of automated crash detection and repair techniques in notebook-based ML development.

\section{Threats to validity}
\paragraph{Dataset representativeness.}
JunoBench contains 111 reproducible crashes, with approximately 15-20 cases for each major ML library and notebook-specific issue. 
During construction, we explicitly aimed to balance coverage across common ML libraries, which does not imply that \emph{all} ML architectures and use cases are represented.  

The benchmark does not aim to mirror the empirical frequency distribution of crashes in ML notebooks observed in the wild.
Specifically, the dataset focuses on ML-related crashes occurring within Python notebooks. 
We intentionally exclude general Python bugs unrelated to ML libraries, as well as hardware- and environment-level crashes, since such issues are difficult to deterministically reproduce and archive. 
Although our post hoc analysis shows that JunoBench includes the major crash types and root-cause categories identified in a prior empirical study of ML notebook crashes~\cite{wang2025machinelearningnotebookscrash}, the moderate overall size of the benchmark implies that less common crash types and edge-case API behaviors are likely to be underrepresented. 

Furthermore, we focus on crashes that interrupt notebook execution (e.g., uncaught exceptions or runtime errors), thereby, excluding non-crashing defects such as silent logic errors, performance degradation, or suboptimal modeling behavior.
Moreover, the dataset is derived exclusively from publicly available Kaggle notebooks that facilitate crash reproduction. 
While Kaggle reflects common ML development practices, it may not capture industrial workflows. 
Consequently, certain crash types common in industrial environments may be underrepresented.

Therefore, JunoBench can serve as a curated and reproducible benchmark for evaluation purposes rather than as a comprehensive census of all possible ML notebook bugs.

\paragraph{Potential annotation bias.}
All crash categorizations and diagnosis labels were manually assigned, which may introduce subjective bias. 
To mitigate potential bias, two annotators were involved in the labeling process, and disagreements were resolved by discussion. 
For categorical labels, Cohen’s $\kappa$ exceeds 0.9 across all dimensions, indicating almost perfect agreement. 
Diagnosis labels were grounded in explicit execution evidence, including crashing code cells, error outputs, and verified fixes, and were validated by a second annotator. 
Despite these measures and practices, some subjectivity may remain, particularly for complex crashes.

\paragraph{Library evolution and environment stability.}
ML libraries evolve over time, with API changes, behavioral modifications, and deprecations occurring across versions. 
JunoBench captures the behavior of library versions available at the time of benchmark construction (June 2025). 
To ensure long-term reproducibility, we archived a unified Docker image with fixed dependency versions, enabling deterministic re-execution of all benchmark cases independent of upstream changes.
However, the benchmark may not reflect crashes introduced in future library releases. 
As APIs evolve, new crash patterns may emerge while older ones may disappear. 
Future extensions of JunoBench will require systematic incorporation of crashes from newer library versions together with corresponding versioned execution environments to preserve reproducibility across releases.

\section{Conclusion}
This paper presents JunoBench, a benchmark for studying crash detection, diagnosis, and repair in real-world ML notebooks. JunoBench contains 111 curated and reproducible crashes in Python-based ML notebooks, each paired with a verifiable fix, enabling controlled and reliable evaluation of debugging techniques under realistic notebook execution behavior.

The benchmark covers major ML libraries, including \textit{TensorFlow/Keras}, \textit{PyTorch}, \textit{Scikit-learn}, \textit{Pandas}, \textit{NumPy}, \textit{Matplotlib}, and \textit{Seaborn}, as well as notebook-specific issues such as out-of-order execution. Each crash is accompanied by human-validated categorizations, including \emph{library cause}, \emph{crash type}, \emph{root cause}, and \emph{ML pipeline stage}, together with ground-truth labels for crash detection and diagnosis. In addition, JunoBench provides a unified execution environment that ensures consistent and reproducible evaluation across all benchmark cases.

By combining reproducible crashes, verified fixes, structured annotations, and a reliable execution environment, JunoBench supports a broad range of research directions, including execution-aware debugging, pipeline-aware and cross-library crash analysis, interactive debugging agents, and LLM-based crash reasoning. 

\section{Dataset availability}

JunoBench is available on HuggingFace~\cite{junobenchrepo}, and all artifacts used to construct the benchmark are provided on GitHub~\cite{wang2025code}.


\newpage
\begin{acks}
This work was partially supported by the Wallenberg AI, Autonomous Systems and Software Program (WASP) funded by the Knut and Alice Wallenberg Foundation, and the Software Center Project 61.
\end{acks}

\balance
\bibliographystyle{ACM-Reference-Format}
\bibliography{References}

\end{document}